\documentclass[aps,prl, twocolumn,groupedaddress,nofootinbib,notitlepage,showpacs,floatfix]{revtex4-1}

\usepackage{graphicx,graphics,epsfig,subfigure,times,bm,bbm,amssymb,amsmath,amsfonts,amsthm,mathrsfs,MnSymbol}
\usepackage[matrix,frame,arrow]{xypic}
\usepackage[pdfstartview=FitH]{hyperref}
\usepackage{pifont}
\hypersetup{
    colorlinks=true,       
    linkcolor=red,          
   citecolor=magenta,        
    filecolor=magenta,      
    urlcolor=cyan,           
    runcolor=cyan
}
\usepackage[pdftex]{color}
\usepackage{braket}
\usepackage{enumerate}
\usepackage[normalem]{ulem}

\usepackage{subfigure}
\usepackage{overpic}

\usepackage{multirow}
\newcommand{\be}{\begin{equation}}
\newcommand{\ee}{\end{equation}}
\newcommand{\ba}{\begin{eqnarray}}
\newcommand{\ea}{\end{eqnarray}}

\newcommand{\ignore}[1]{}

\begin{document}

\title{Many-Body Localization Transition, Temporal Fluctuations of the Loschmidt Echo, and Scrambling}

\author{Jun Yang}
\affiliation{Center for Quantum Information, Institute for Interdisciplinary Information Sciences, Tsinghua University, Beijing 100084, P.R. China}

\author{Alioscia Hamma }
\affiliation{Department of Physics, University of Massachusetts Boston, 100 Morrissey Blvd, Boston MA 02125}
\begin{abstract}
We show that the transition between a ETH phase and a many-body localized phase is marked by the different finite size scaling behaviour of the decay of the Loschmidt Echo and its temporal fluctuations - after a quantum quench - in the infinite time limit, despite the fact that the finite time behaviour of such quantities is dramatically different approach the MBL phase, so that temporal fluctuations cannot be inferred from the infinite time average of the Loschmidt Echo. We also show the different scrambling powers of ETH and MBL Hamiltonians as a probe to the different approaches to equilibrium.
\end{abstract}

\maketitle

\ignore{
\begin{figure}[t!]
\centering
\includegraphics[width=1\columnwidth,clip=true]{concurrence_linear}
\caption{Concurrence as a function of time between spins at distance
  $r$ for the model in the ergodic phase. All the concurrences decay
  to zero by $t=5$, showing that the entanglement is all block
  entanglement.}
\label{concurrence} 
\end{figure}
}

{\em Introduction.---}
Many-body localization (MBL) is a  phenomenon occurring in some interacting quantum many-body systems in presence of a critical value of strong disorder which drives the transition from a generic thermalizing phase. In recent years, MBL has received an enormous attention
\cite{huse-mbl, PhysRevB.91.081103,PhysRevB.92.220201,PhysRevB.88.014206, abanin1, abanin2, abaninlog, oganesyan07, xxz-affleck, scardicchio2} due to its extremely rich and novel phenomenology. MBL is characterized by a lack of thermalization and even equilibration, logarithmic growth of entanglement, logarithmic light cone for the spreading of correlations, area law for the entanglement in high-energy eigenstates, quasi-local conserved quantities of motion, without being integrable systems. Indeed, integrable systems do equilibrate, typically to the generalized Gibbs ensemble, whereas the slow dynamics induced by MBL Hamiltonians keeps the system away from equilibrium.

In this paper we set out to describe the behavior of MBL systems away from equilibrium after a quantum quench by studying the temporal fluctuations of observables. Equilibration is indeed ensued when such fluctuations vanish in the thermodynamic limit. The amount of fluctuations in any observable is ruled by the amount of revivals in the wave-function, that is measured by the Loschmidt Echo (LE)\cite{ref11,zanardiLE}, namely the overlap of the initial wave-function with that at a later time. This overlap is part of a number of fidelity measures that have been capable of detecting and study in detail quantum phase transitions. We show that the MBL transition being mainly a dynamical phenomenon, is well understood by LE and fidelity measures, and that the critical value for disorder can be detected by such measures. 
To this end, we study  the away from equilibrium behavior of the disordered Heisenberg spin-1/2 chain after a quantum quench.  In the weak disorder regime, the system is non-integrable and it does thermalize through the mechanism of eigenstate
thermalization hypothesis' (ETH)\cite{eth1, eth2}, namely the fact that the highly excited states of the Hamiltonian yield thermal expectation values for most local observables. Increasing disorder will not make the system integrable, however, for a critical value of the disorder a transition happens to MBL. The ETH hypothesis breaks down, and even the diagonal ensemble will not return thermal expectation values. This ensemble should in any case reproduce the behaviour of observables in the long time regime after (non thermal) equilibration has happened. However, if local observables have strong time fluctuations, this means that even locally the system is still away from equilibrium. In order to study this behaviour, we resort to fidelity. We define fidelity between two quantum states  $\rho$ and $\sigma$ as 
$\mathcal{F}(\rho, \sigma) = \mathrm{Tr} \sqrt{\rho^{1/2}\sigma\rho^{1/2}}$ which reduces to the overlap of wave functions in the case of pure quantum states $\rho = \ket{\psi}\bra{\psi}, \sigma = \ket{\varphi}\bra{\varphi}$.
Previous work \cite{PhysRevB.92.014208,PhysRevB.92.220201} has investigated the MBL transition using fidelity and Loschimidt Echo (LE). In \cite{PhysRevB.92.220201} the fidelity between two ground states shows that both Anderson localization and MBL feature orthogonality catastrophe in the ground state. In \cite{PhysRevB.92.014208}, it is shown a power-law decay with time for the LE after a sudden quench with a completely factorized state as initial state. In this work, we are interested in locating the MBL transition point and describing the fate of temporal fluctuations over long times. We take seriously the idea that the slow dynamics is the signature of the dynamical phase transition to MBL and thus expect the transition to be marked by the different long time fluctuations. 

The Loschmidt Echo $\mathcal{L}(t)$ is defined as the  square of the fidelity $\mathcal{F}$ between the initial state $\rho(0)$ and its time evolution $\rho(t)$, for pure states, $\mathcal{F}(\ket{\psi(0)},\ket{\psi(t)}) = |\braket{\psi(0)|\psi(t)}|$, which is the square root of $\mathcal{L}(t)$. This quantity is involved in Equilibration because of the inverse participation ratio $\mathrm{IPR}_q $. Indeed, given a 
 wave function $\ket{\psi} = \sum_n C_n\ket{n}$,  the  inverse participation ratio is defined by $\mathrm{IPR}_q = \sum_n |C_n|^{2q}$\cite{PhysRevB.92.014208}. For  $q = 2$ this is just the infinite time average of LE\cite{zanardiLE}. In the Anderson localization transition, the average over realizations of $\mathrm{IPR}_q$ scales\cite{PhysRevB.92.014208}\cite{RevModPhys.80.1355}\cite{PhysRevB.64.241303}
\begin{equation}
	\braket{\mathrm{IPR_q}} \sim \mathcal{D}^{-(q-1)D_q}
\end{equation}
where $\mathcal{D}$ is the dimension of the space we focus on.  $D_q$ is an indicator determining whether or not the state is localized. In the localized phase $D_q = 0$, in the ergodic phase $D_q = d$, where $d$ is the system dimension. 

The infinite time average of LE gives an upper bound for the temporal fluctuation of an observable $A$. Under the non-resonant assumption, which says $E_n-E_m=E_k-E_l$ if and only if $n=m, k = l$ or $n=k, m=l$, the following result holds\cite{doi:10.1142/9789814704090_0008}\cite{PhysRevLett.101.190403}
\begin{equation}
	\Delta A^2 \leq \|A\|^2 \overline{\mathcal{L}(t)}^\infty =\|A\|^2 \cdot\mathrm{IPR_2}
	\label{upperbound}
\end{equation}
As the norm of a local observable is $O(1)$,  a small $\mathrm{IPR_2}$ compresses the temporal fluctuation of a local observable. In the Anderson localization situation, the $\mathrm{IPR_2}\rightarrow 0$ when $L\rightarrow \infty$ in the thermalized phase. But the localized phase gives $D_q = 0$, resulting in a constant $\mathrm{IPR_2}$, a local observable still fluctuates in the thermodynamical limit $L\rightarrow \infty$.

We examine the size scaling of $\mathrm{IPR}_q$ in the MBL phase transition. The $\mathrm{IPR}_q$ is computed from an initial state being an exact eigenstate of $H$ in the middle of the spectrum, while its participation is computed in the basis of the eigenstates of a Hamiltonian $H'$ where the couplings $h_i$ have been quenched to a random slightly different value. So this is the scenario of a small quantum quench.   The size scaling shows similar behavior with the Anderson localization transition, with large $D_q$ in the ETH phase and small $D_q$ in the MBL phase. The difference in $D_q$ between the two phase shows the phase transition critical point.
Notice that these fluctuations are bound in the limit of infinite time average. As we shall see, in this limit LE decays exponentially with the system size $L$ both in the ETH and the MBL phase, resulting in zero fluctuations. However, here the infinite time limit is misleading. We need to consider finite times $T$ to average over, and finite sizes $L$. We show that  the scaling of the fluctuations with both $L,T$ results in a fast equilibration for the ETH phase and a survival of time fluctuations for the MBL phase. In other words, the order of limits counts and if one takes first the thermodynamic limit the MBL dynamics is non-equilibrating. 

{\em Model.---}
The Hamiltonian for the isotropic Heisenberg spin-1/2 model is
\begin{equation}
H=\sum_{i=1}^L J\left(\vec{\sigma}_i\cdot \vec{\sigma}_{i+1}\right)+\Gamma \sigma_i^x + h_i \sigma_i^z
\label{H}
\end{equation}
We set $J=1$. We introduce a small constant field $\Gamma = 0.1$ along the $x$ direction to break the conservation of the total angular momentum $S^z$\cite{PhysRevLett.115.267206}.  Disorder is in the random couplings $h_i$ indipendently uniformly distributed in the interval $[-h,h]$. Periodic boundary conditions are assumed in the following unless specifically notified. If the total angular momentum $S_z$ is conserved, in the $S_z = 0$ sector the systems featuers MBL transition at $h=h_c\approx 3.5\pm1.0$\cite{PhysRevB.82.174411}. Breaking  $S_z$ conservation by $\Gamma$ results in a critical point  $h_c \approx 3.3$\cite{PhysRevLett.115.267206}.

We compute by exact diagonalization the behavior of fidelity the initial state and the state at time $t$ by quenching the Hamiltonian from $H(\vec{h})$ to $H(\vec{h^\prime})$, where $\vec{h^\prime}=\vec{h}+\vec{\delta h_i}$ and $\vec{\delta h_i}$ is a random  vector whose components lie in the small interva $[-0.1,0.1]$.
In our scheme, the initial state $\ket{\psi_0}$ is chosen to be the highly excited eigenstate of Hamiltonian $H(\vec{h})$, with energy $0$, which is exactly in the middle of the energy spectrum  in this model. 

 \begin{figure}[h]
\centering
\includegraphics[width=0.7\linewidth]{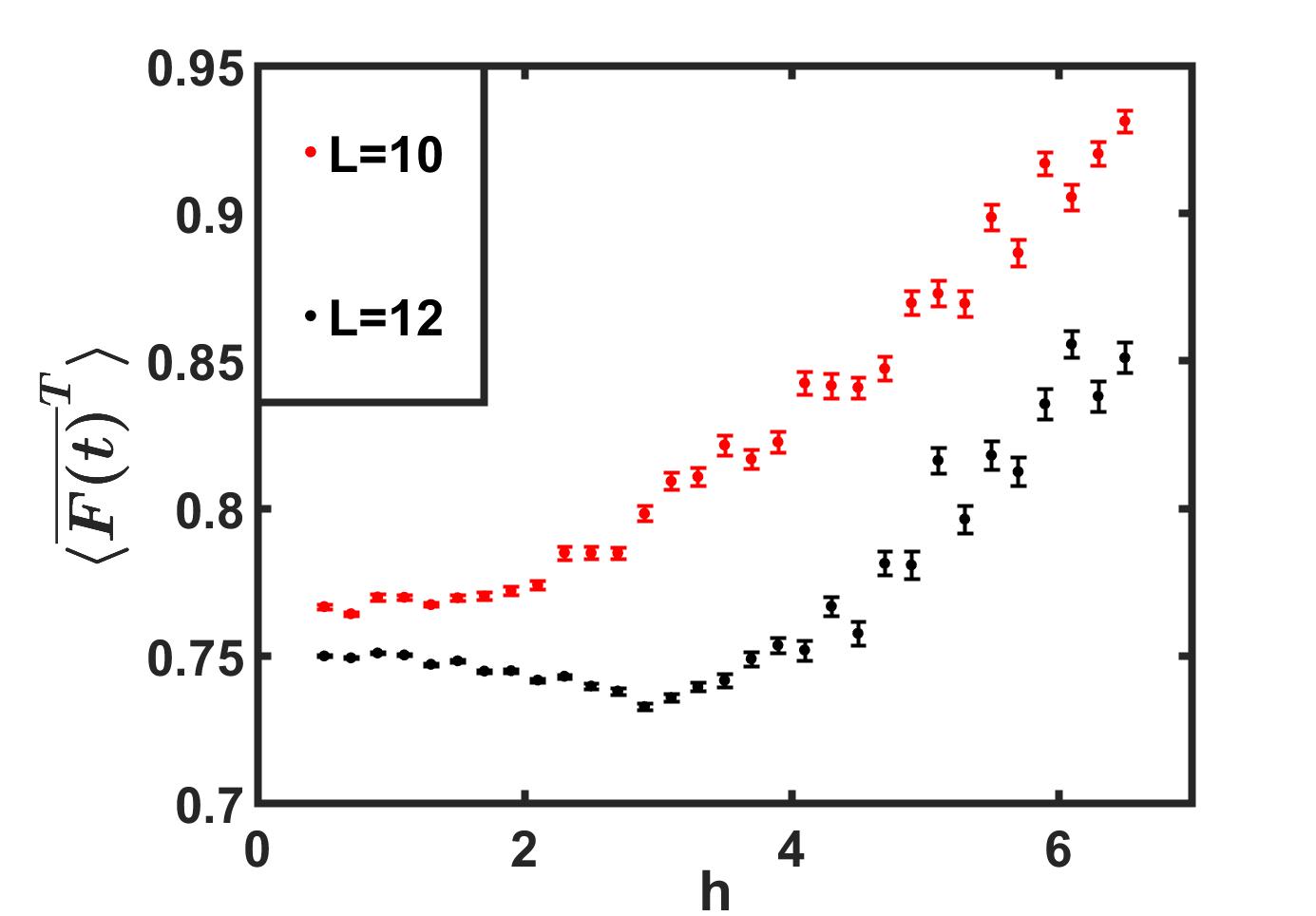}
\caption{Time average of subsystem fidelity with different initial disorder  $h$. Red dots for $L = 10$, black dots for $L = 12$. In both cases, the data points are averaged over 500 realizations. The time interval is $[0,10^8]$}
\label{fig:fidAvg}
\end{figure}

\begin{figure}[h]
	\subfigure[]{
		\includegraphics[width=0.45\linewidth,height=0.4\linewidth]{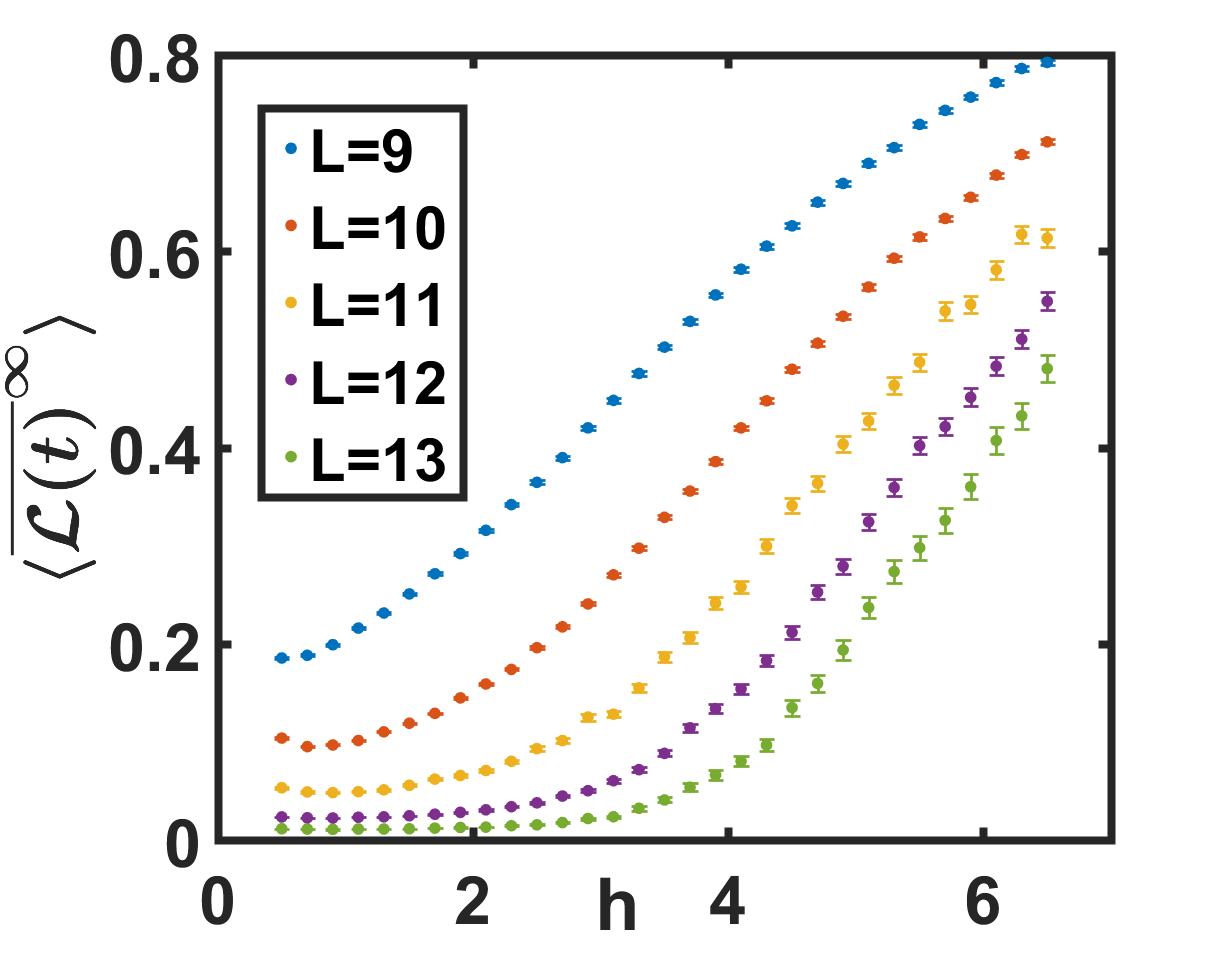}
		\label{fig:LEAvg}
	}
	\subfigure[]{
		\includegraphics[width=0.45\linewidth,height=0.4\linewidth]{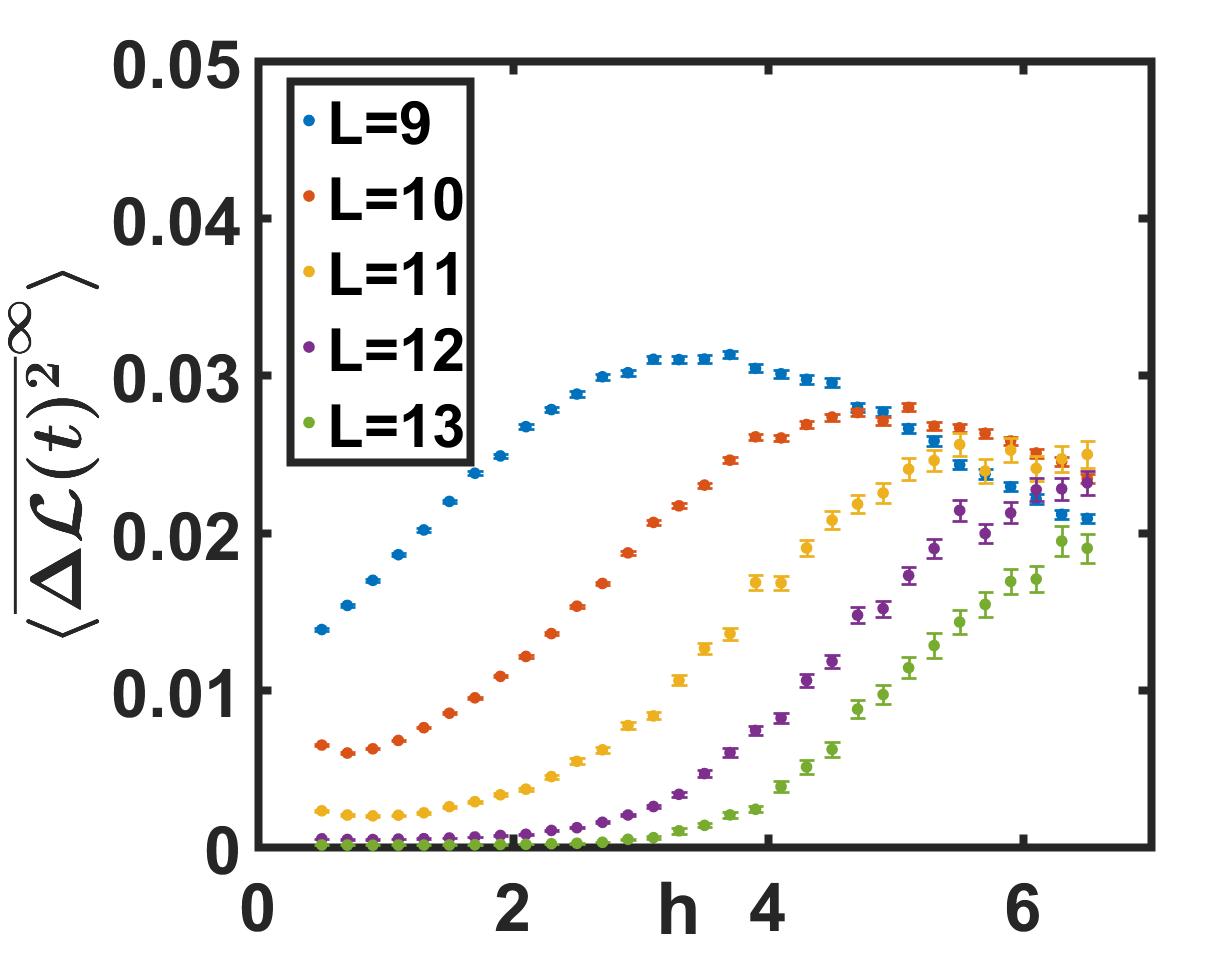}
		\label{fig:LEVar}
	}
	\caption{(a)Infinite temporal average of Loschmidt Echo with different initial $h$ for system size $L=9\sim13$. (b)Temporal fluctuation of Loschmidt Echo with different initial $h$  for system size $L=9\sim13$ }
	
\end{figure}

{\em Fidelity in the infinite time limit.--} Here, we want to show that the study of LE and its temporal fluctuations in the infinite time limit marks the MBL transition.
We compute the fidelity both as an overlap of the total wave-function $\psi(t)$ with the initial state $\psi(0)$, but also as the overlap of the marginal states on a subsystem. The subsystem is defined as follows:  we choose the last $m$ spin sites of the total $L$ spin sites as our subsystem, which is labeled as $B$, the remaining part is labeled as $A$, thus the state of the subsystem $B$ is $\rho_B(t) = \mathrm{Tr}_A (\ket{\psi(t)}\bra{\psi(t)}) $. 
The fidelity between the $\rho_B(0)$ and $\rho_B(t)$ then reads $\mathcal{F}(t)= \mathrm{Tr} \sqrt{\rho_B(0)^{1/2}\rho_B(t)\rho_B(0)^{1/2}}$. For every value of $h$, we calculate its time average over a time $T=10^8$ for each realization and then average it over different realizations. In Fig:\ref{fig:fidAvg} we have chosen $m=L/2$ and plotted the behaviour of the average fidelity with $h$. The behaviors with the disorder strength are completely different in the two sides of the critical point $h_c \approx 3.1$. It keeps almost invariant when the disorder strength is smaller than $h_c$, but then it grows with  the disorder strength past the critical point $h_c$. As we can see from Fig.\ref{fig:fidAvg}, however, the value $h_c \approx 3.1$  cannot detected accurately. This is due to small subsystem size. It is anyway interesting to see that a qualitative different behavior can be seen on the subsystem overlap. So let us move to consider the overlap of the full wave function, namely the Loschmidt Echo 
 \begin{equation}
	 \begin{split}
 	 \mathcal{L}(t) &= |\braket{\psi(0)|e^{-iH(h^\prime)t}|\psi(0)}|^2\\ &= \left| \sum_n |C_n|^2 e^{-iE_n t}\right| ^2 \\
 	 &= \sum_n |C_n|^4 + \sum_{n\neq m}|C_n|^2|C_m|^2e^{-i(E_n-E_m)t}
 	 \label{LE}
 	 \end{split} 
 \end{equation}
where $C_\alpha = \braket{\psi(0)|n}$, $\ket{n}$ is the $n$-th eigenstate of $H(h^\prime)$. 
Because of disorder, the assumption of the non-resonant condition holds\cite{doi:10.1142/9789814704090_0008}. and  the infinite time average of LE  becomes
\begin{equation}
	\overline{\mathcal{L}(t)}^\infty = \sum_n |C_n|^4
\end{equation}
Moreover, the infinite time average of temporal fluctuations reads
\begin{equation}
\begin{split}
\overline{\Delta\mathcal{L}(t)^2 }^\infty&= \overline{\left(\mathcal{L}(t)-\overline{\mathcal{L}(t)}^\infty \right)^2}^\infty\\&=\left(\sum_n |C_n|^4\right)^2-\sum_n |C_n|^8 
\label{varLE}
\end{split}
\end{equation}
We compute these quantities as a function of the disorder strentgh $h$ and see that they both show different behaviors in the ETH phase and MBL phase. Fig2:\ref{fig:LEAvg} shows the infinite temporal average of the Loschmidt Echo.  In the ETH phase the average LE is small and independent of disorder strength. However, at the transition is increases dramatically and then keeps growing with disorder strength in the MBL phase. In panel Fig2:\ref{varLE}, the temporal fluctuations show a similar behavior. A critical point at $h_c\approx 3.1\pm.3$ can be located to mark the phase transition. The data points are averaged over 10000 realizations for $L = 9$, 10000 realizations for $L =10$, 1000 realizations for $L =11,12$, 500 realizations for $L=13$.

\begin{figure}[h]
	\subfigure[]{
		\includegraphics[width=0.45\linewidth,height=0.4\linewidth]{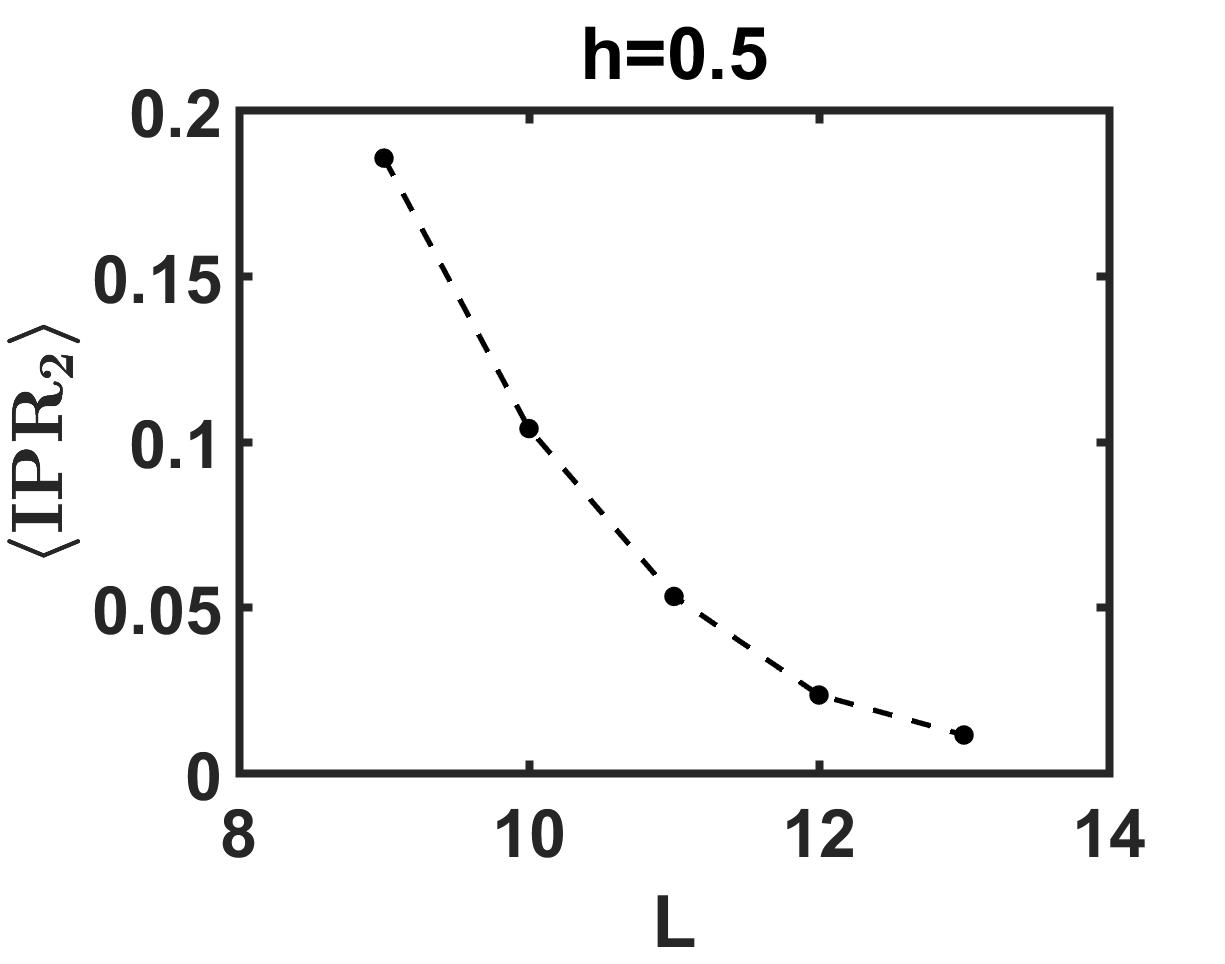}
		\label{fig:LE0_5}
	}
	\subfigure[]{
		\includegraphics[width=0.45\linewidth,height=0.4\linewidth]{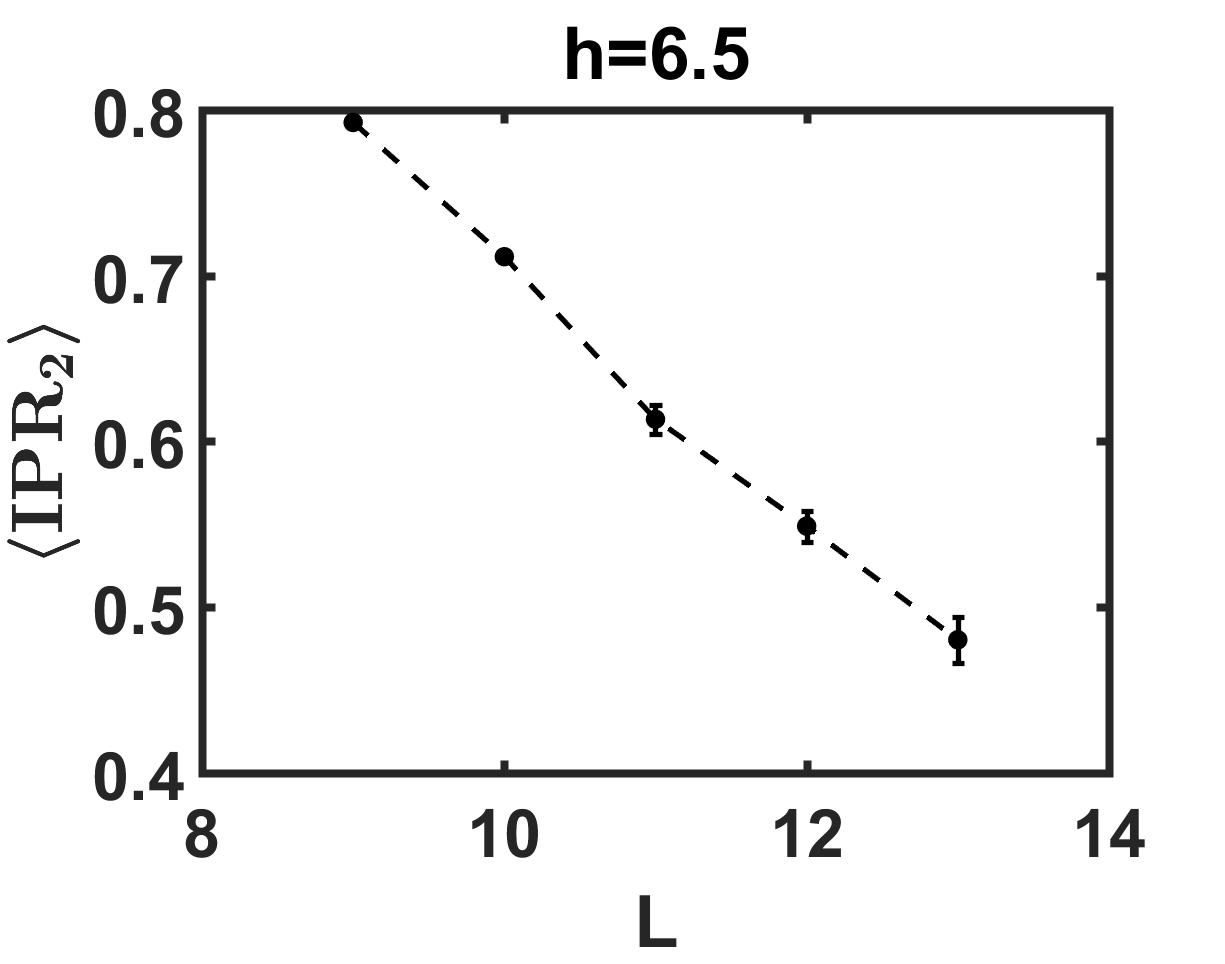}
		\label{fig:LE6_5}
	}
	\caption{$\mathrm{IPR_2}$ scaling with the system size. (a) $\braket{\mathrm{IPR_2}}$ vs system size $L$ for $h=0.5$ (b) $\braket{\mathrm{IPR_2}}$ vs $L$ for $h=6.5$,}
\end{figure}

\begin{figure}[h]
		\subfigure[]{
			\includegraphics[width=0.45\linewidth,height=0.4\linewidth]{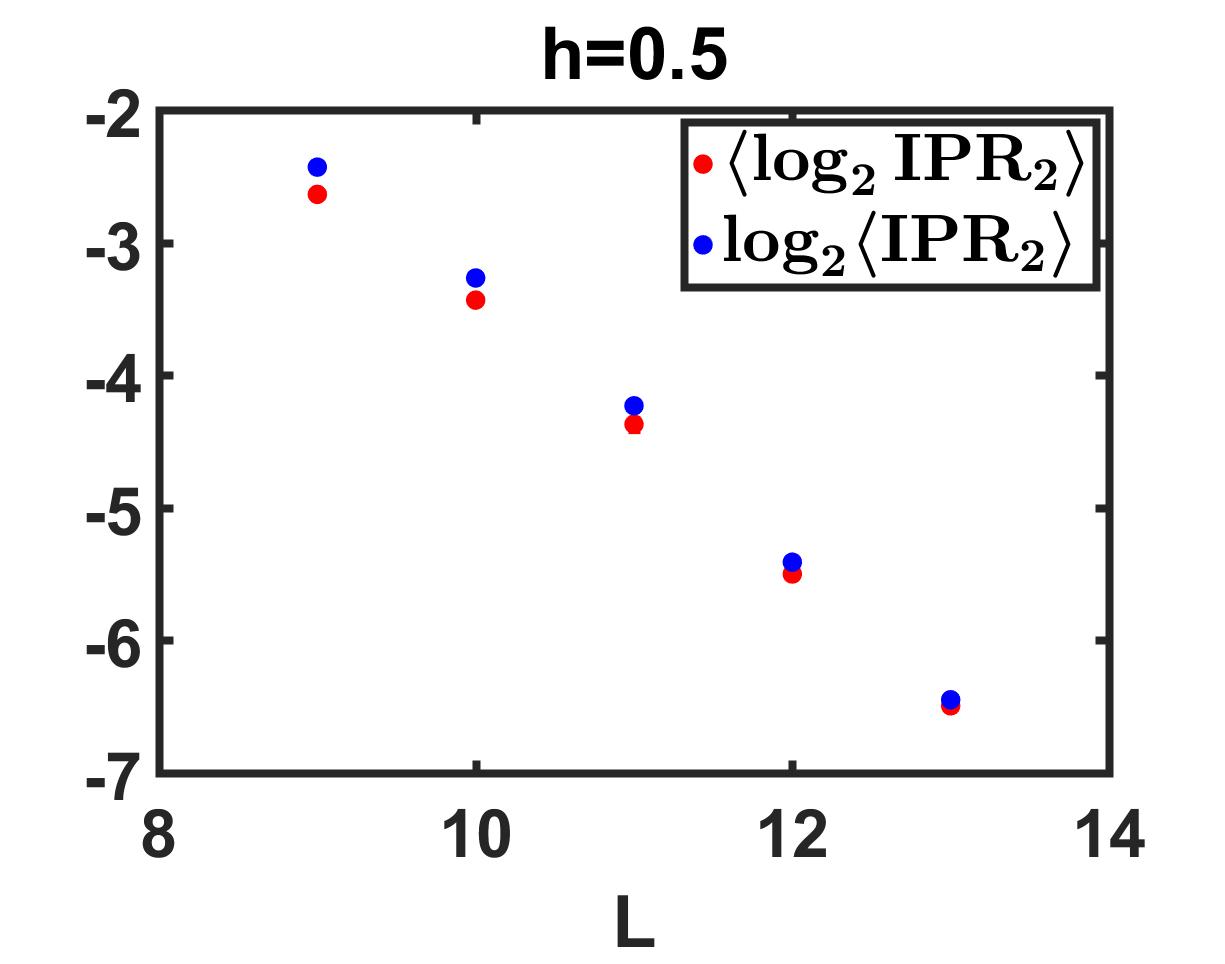}
			\label{fig:logLE0_5}
		}
		\subfigure[]{
			\includegraphics[width=0.45\linewidth,height=0.4\linewidth]{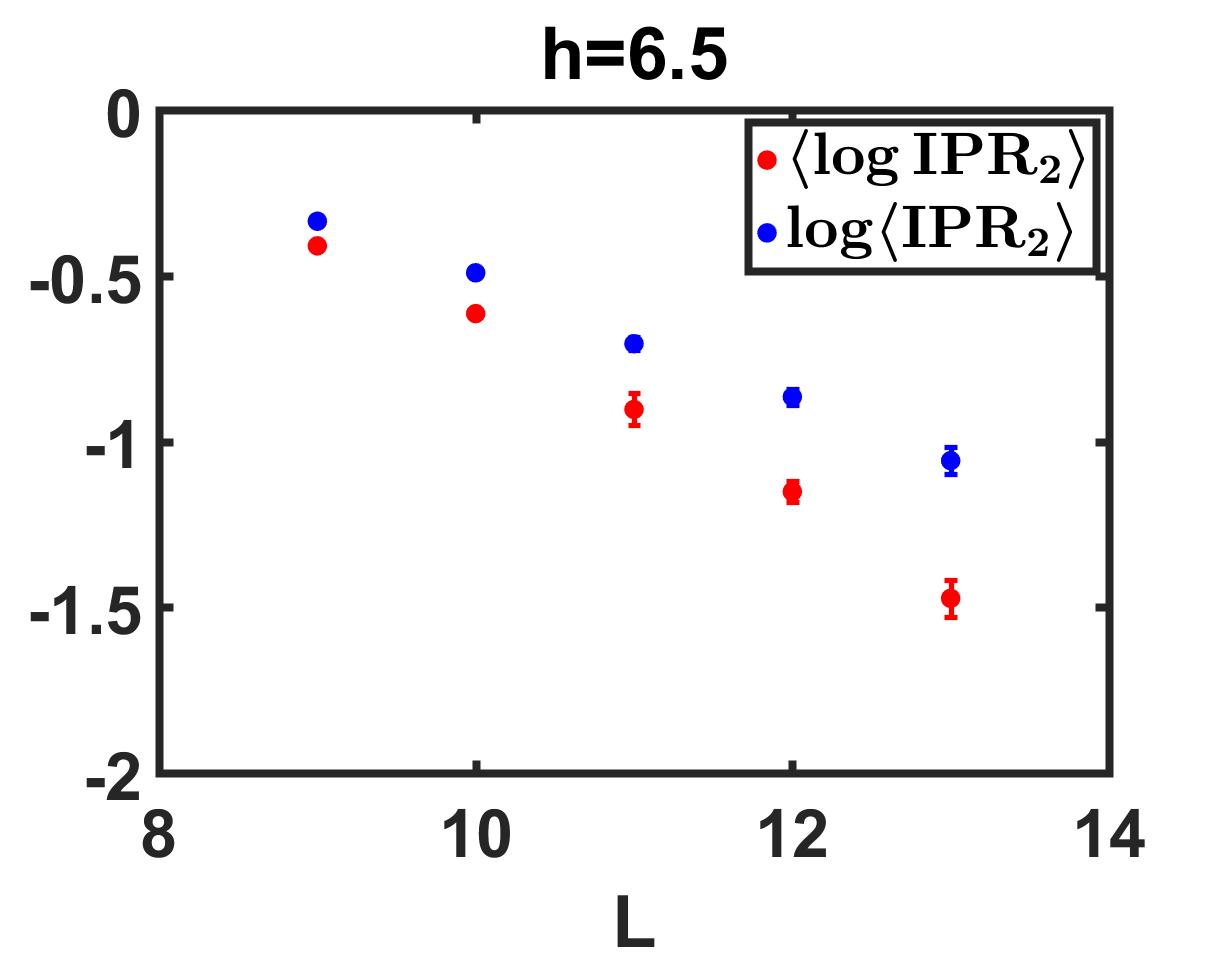}
			\label{fig:logLE6_5}
		}	
			\caption{(a)Blue dots: $\log_2\braket{\mathrm{IPR_2}}$ vs $L$ for $h=0.5$. Red dots: $\braket{\log_2(\mathrm{IPR_2})}$ vs $L$ for $h=0.5$.(b) 
				Blue dots: $\log_2\braket{\mathrm{IPR_2}}$ vs $L$ for $h=6.5$. Red dots: $\braket{\log_2(\mathrm{IPR_2})}$ vs $L$ for $h=6.5$.}
\end{figure}

\begin{figure}[h]
 	\subfigure[]{
 		\includegraphics[width=0.45\linewidth,height=0.4\linewidth]{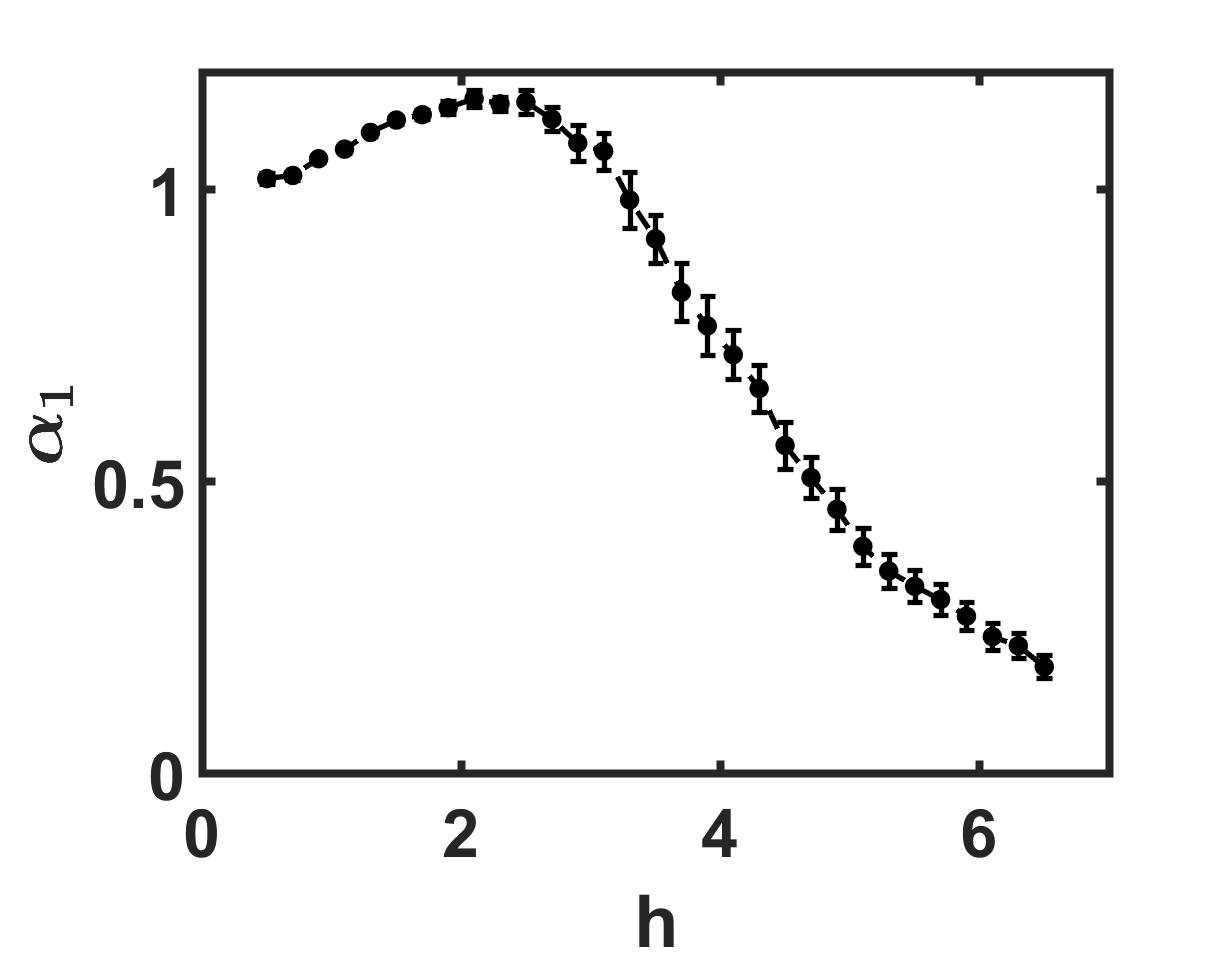}
 		\label{expdecay1}
 	}
 	\subfigure[]{
 		\includegraphics[width=0.45\linewidth,height=0.4\linewidth]{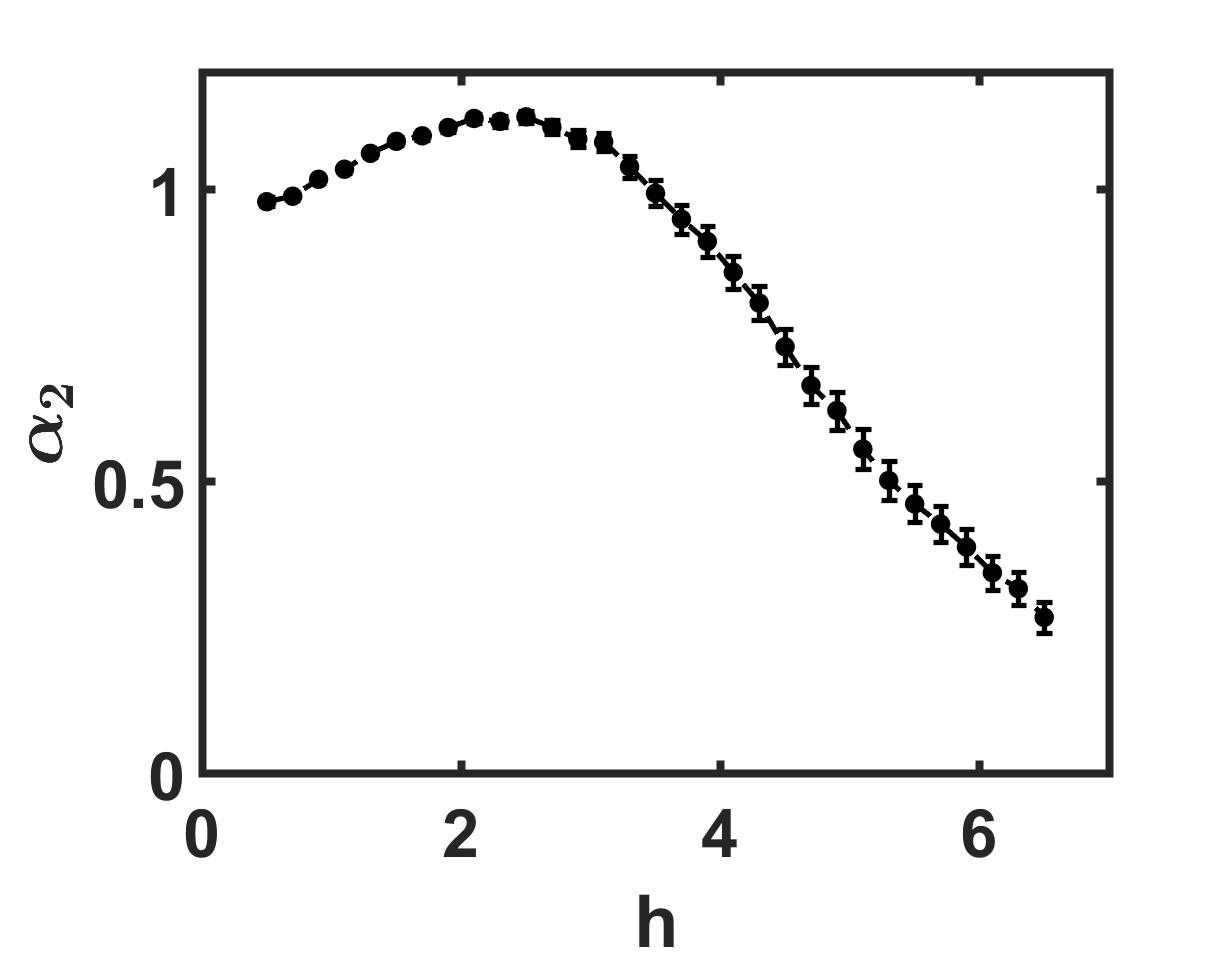}
 		\label{expdecay2}
 	}	
 	\caption{(a)$\alpha_1$ vs disorder strength $h$. (b)$\alpha_2$ vs disorder strength $h$ all the data point are realized over 500 realizations for each $h$ for $L = 13$, 1000 realizations for $L = 11,12$, 10000 realizations for $L = 9,10$}
 \end{figure}

At this point, we are interested in how the above quantities scale with system size deep in the ETH and MBL phases. In Fig:(3) we show $\braket{\mathrm{IPR_2}}$ and  $\log_2\braket{\mathrm{IPR_2}}$ scaling with the system size $L$. 
In Fig:\ref{fig:LE0_5}, we see the finite size scaling  in the ETH phase($h=0.5 < h_c$) while Fig:\ref{fig:LE6_5} shows the $\braket{\mathrm{IPR_2}}$ scaling with the system size in the MBL phase($h=6.5 > h_c$).
The logarithmic quantities allow us to find a different scaling behavior. We are interested in both the average $\braket{\mathrm{IPR_2}}$ in logarithmic scale and the scaling behavior of $\braket{\log_2(\mathrm{IPR_2})}$. In the ETH phase, with $h=0.5$ we observe an exponential decay of these quantities. Referring to
 Fig:\ref{fig:logLE0_5}, blue dots represent the logarithm  of $\braket{\mathrm{IPR_2}}$ for different system sizes $L$ while red dots show the scaling behavior with $L$ of $\braket{\log_2(\mathrm{IPR_2})}$. As the figure shows, the two lines almost coincide with each other, and the difference get smaller when the system size grows.
Also the MBL phase features an exponential decay, though with a different exponent.  Fig:\ref{fig:logLE6_5} shows the  $\log_2\braket{\mathrm{IPR_2}}$  and the $\braket{\log_2(\mathrm{IPR_2})}$ scaling behavior in the MBL phase. As we can see,  exponent  is much smaller than that in the ETH phase. Moreover, the difference between the two quantities is much larger in the MBL phase, and the difference gets larger as the system size increases.
From Fig:\ref{fig:logLE0_5}\ref{fig:logLE6_5}, we find that the $\log_2\braket{\mathrm{IPR_2}}$  and the $\braket{\log_2(\mathrm{IPR_2})}$ decays linearly with the system size, which means the scaling behavior of the $\braket{\mathrm{IPR_2}}$ in MBL phase transition gives similar result in the Anderson transition, which decays exponentially with the system size $L$. By fitting these two lines, we find two decaying exponents $\alpha_1, \alpha_2$, satisfying $\log_2\braket{\mathrm{IPR_2}} \sim -\alpha_1 L$ and $\braket{\log_2(\mathrm{IPR_2})} \sim -\alpha_2 L$. At this point, 
 we investigate the behavior of the exponents $\alpha_1$ and $\alpha_2$  with the strentgh disorder $h$. We see that this marks the MBL transition as well.
Fig:\ref{expdecay1}\ref{expdecay2} shows how $\alpha_1$ and $\alpha_2$ scales with the disorder strength $h$. Both $\alpha_1$ and $\alpha_2$ grows slightly  in the ETH phase, but after a critical point $h_c \approx 2.9\pm.3$,  $\alpha_1$ and $\alpha_2$ drop linearly with the increasing disorder strength. The transition from a slightly growing exponent to a linear decay exponent is an important feature showing the MBL phase transition. In the thermal dynamic limit, this will result in a sigularity at $h_c\approx 2.9$. We note that the critical point is slightly smaller than that obtained  in the previous section. In \cite{scardicchio1}, it is also found that $IPR$ of initial factorized states can detect the transition point. We remark that the (small) quantum quench scenario investigated here is a physical situation more amenable to experimental observation.

\begin{figure}[h]
 	\subfigure[]{
 		\includegraphics[width=0.45\linewidth,height=0.4\linewidth]{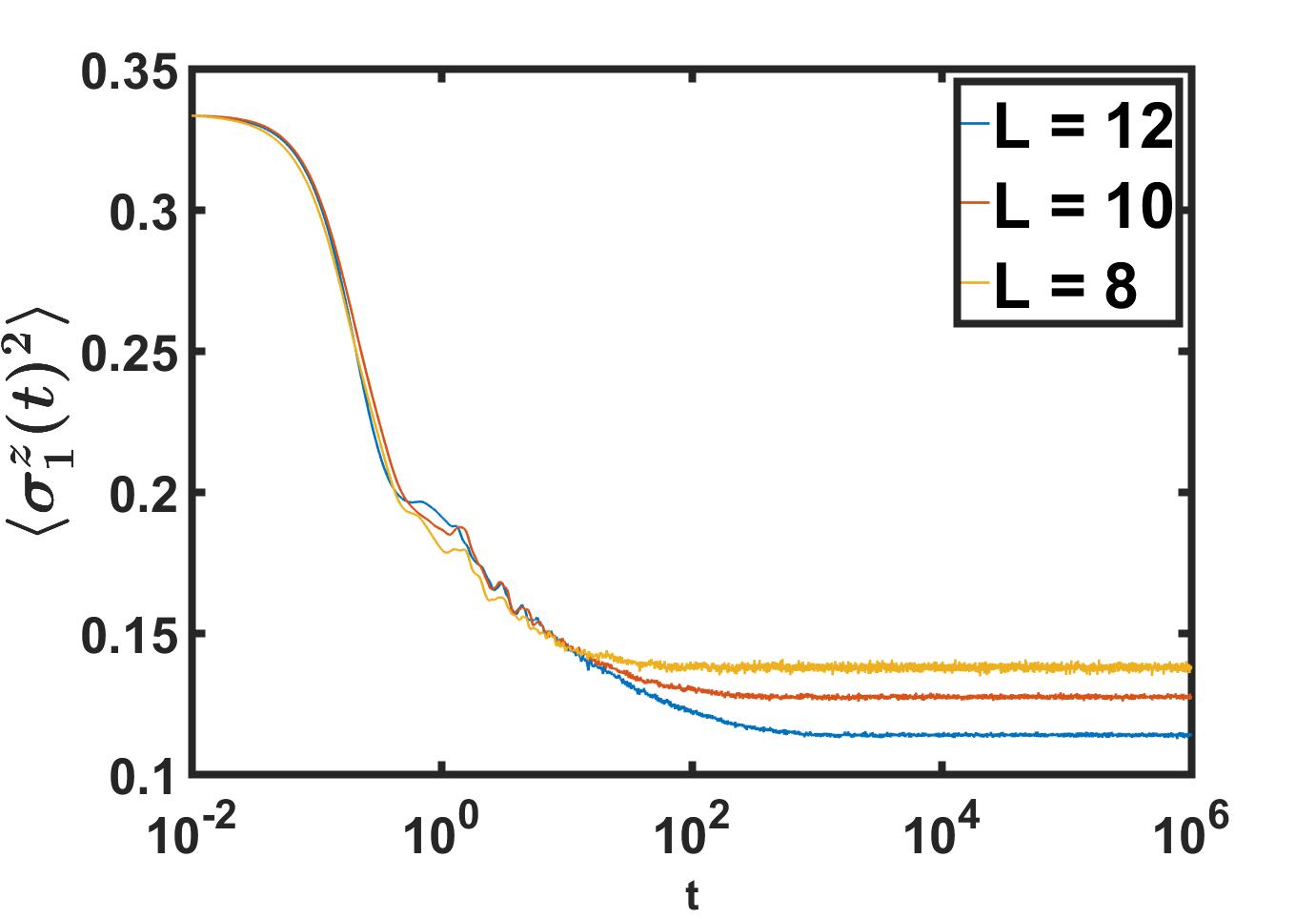}
 		\label{sigma1z_MBL}
 	}
 	\subfigure[]{
 		\includegraphics[width=0.45\linewidth,height=0.4\linewidth]{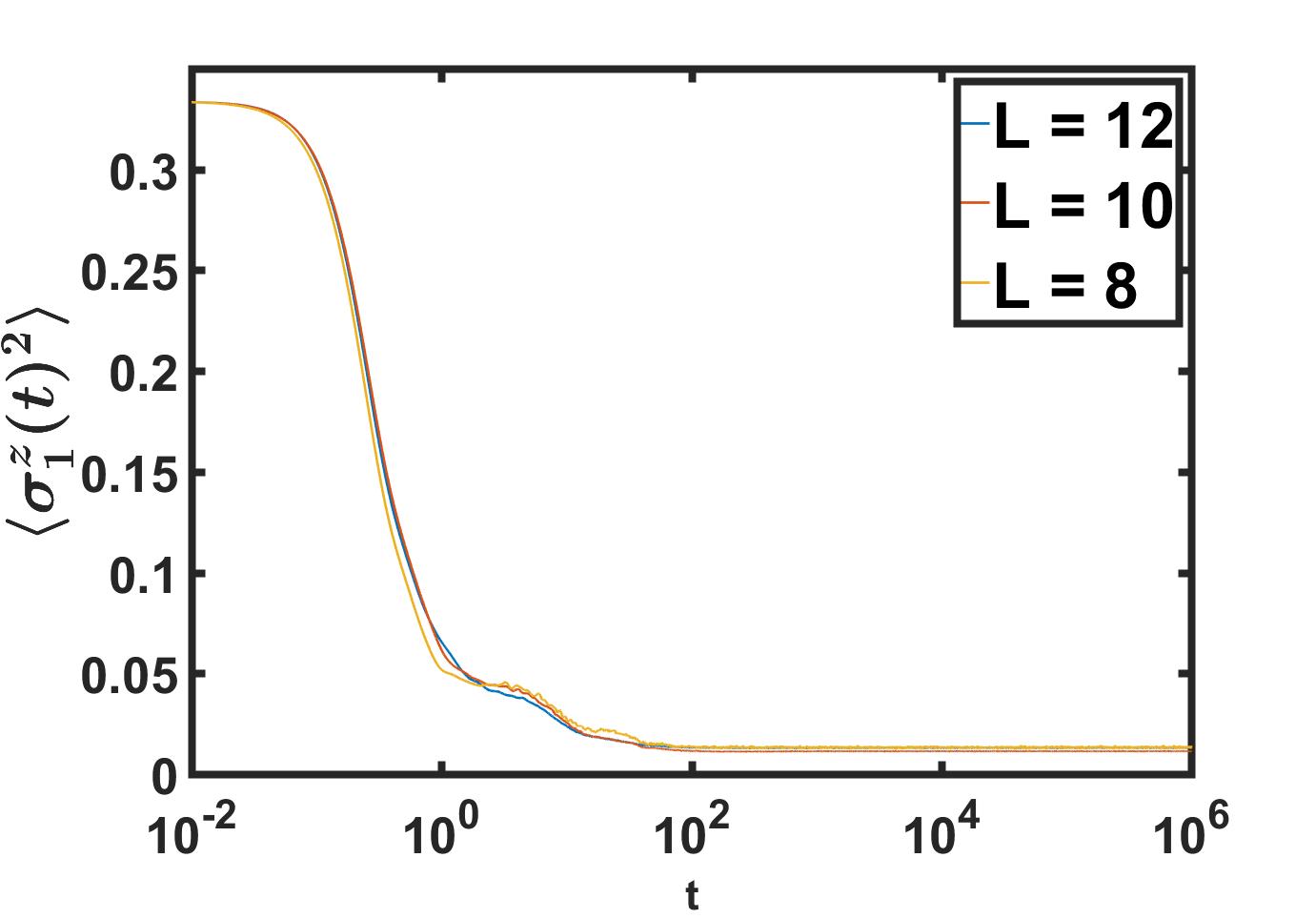}
 		\label{sigma1z_ETH}
 	}	
 	\caption{(a) Time evolution of  $\braket{\sigma_1^z(t)^2}$ in the MBL phase with $h=6.5$ for system size $8,10,12$ (b) Time evolution of  $\braket{\sigma_1^z(t)^2}$ in the ETH phase with $h=0.5$ for system size $8,10,12$.
 	The data are averaged over 1000 realizations for $L = 12$, 2000 realizations for $L=10, 8$.}
 \end{figure}

\begin{figure}[h]
\centering
\includegraphics[width=1\linewidth]{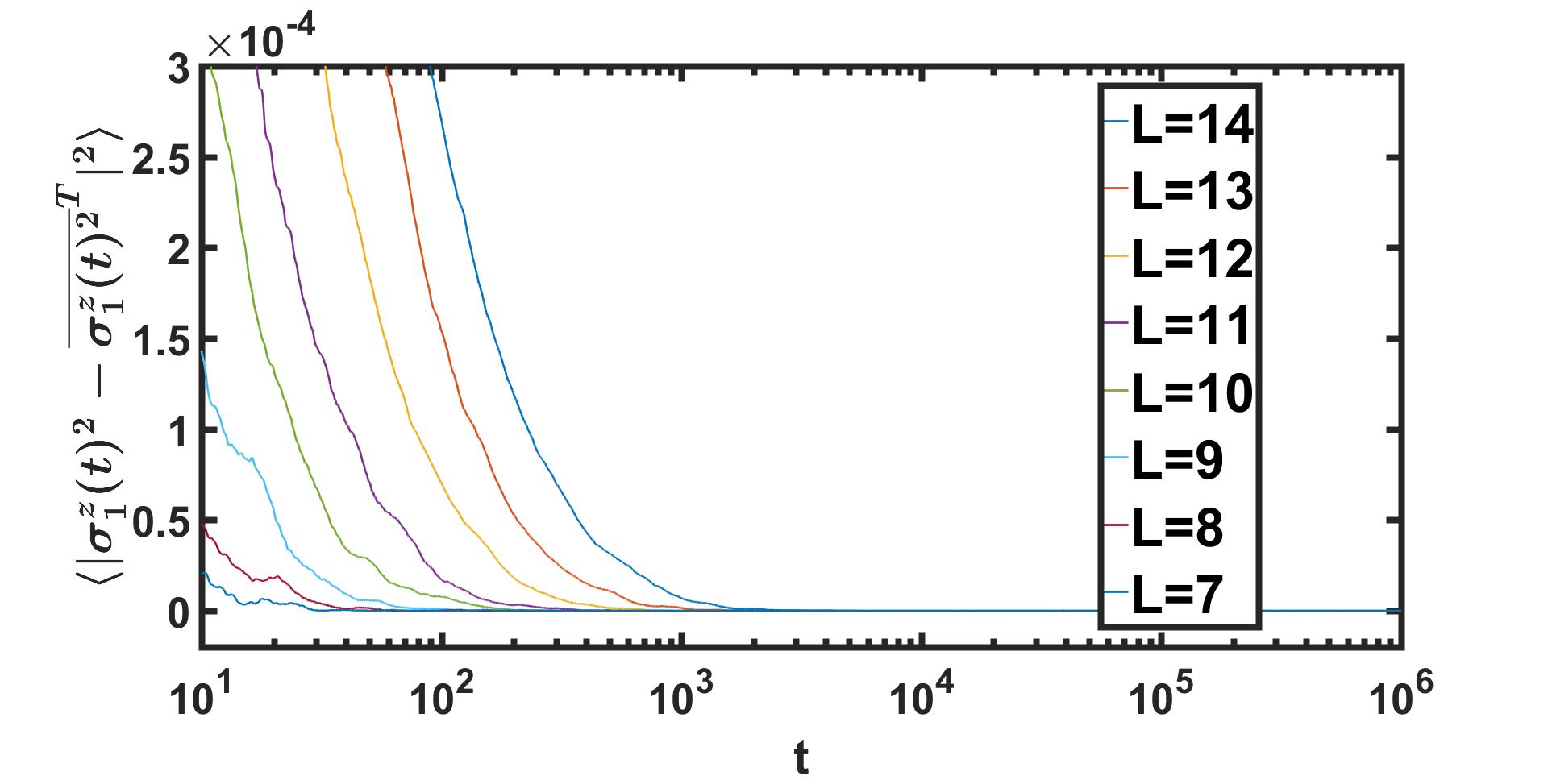}
\caption{Equilibration time scaling with the system size in the MBL phase, the curves are averaged over 2000 realizations for $L = 7-10$ , 1000 realizations for $L = 11, 12$, 500 realzaitons for $L  = 13$, 400 realzaitons for $L = 14$}
\label{EqTMBL}
\end{figure}

\begin{figure}[h]
\centering
\includegraphics[width=1.0\linewidth]{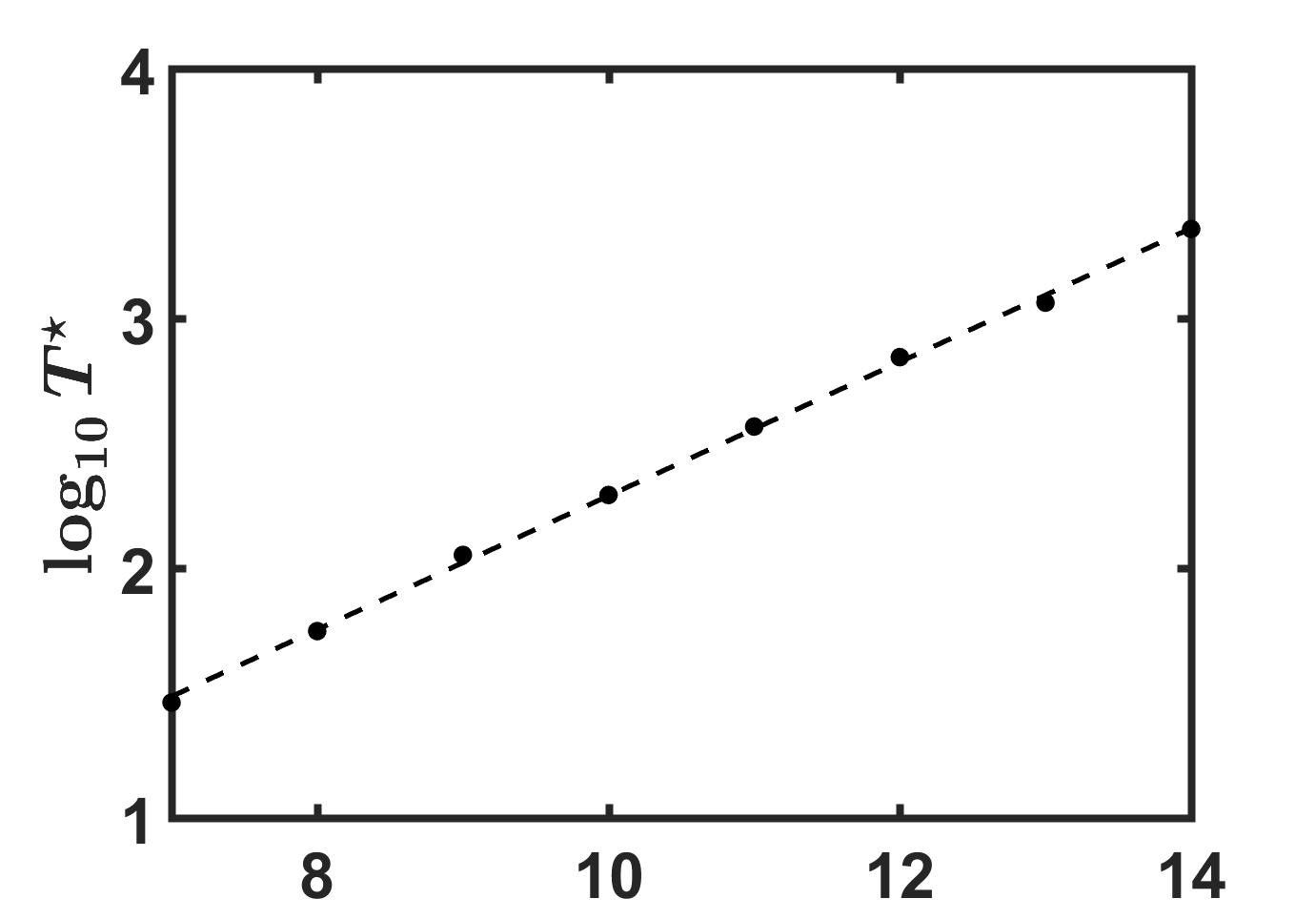}
\caption{scaling behavior of local observable equilibration time $\log(T^\star)$ with the system size $L$ in the MBL phase with $h = 6.5$.}
\label{eqtMBLscaling}
\end{figure}

{\em Equilibration at finite times.---} 
From the observation above, we find the infinite time average of LE decays exponentially with the system size $L$. Although it decays much slower in the MBL phase, the exponential decaying law proves that the infinite temporal fluctuation of a local observable is very small in the thermodynamic limit. However, infinite time can be misleading if the actual time for equilibration grows very fast with system size.   In the following, we show that the equilibration time grows exponentially with the system size in the MBL phase while it is upper bounded by a constant time in the ETH  phase. In this section, we quench from  initial random product states $\ket{\psi(0)} = \otimes_{i=1}^N(\alpha_i\ket{0}_i+\beta_i\ket{1}_i)$. With this setting, we can prove that  $\braket{\mathcal{L}}\leq (2/3)^L$ averaging on infinite time (see Appendix). However, to study the behavior at finite times, let us first focus on the dynamics of the local observable s $\sigma_1^z$. We define $\sigma_1^z(t) = \braket{\psi(t)|\sigma_1^z|\psi(t)}$ and $\braket{\sigma_1^z(t)}$ is the  average of $\sigma_1^z(t)$ over the realizations of initial conditions and Hamiltonians. Here, we average the  $\braket{\sigma_1^z(t)^2}$. A theoretical analysis shows that  $\braket{\sigma_1^z(0)^2}=1/3$ (see Appendix). So we shift the initial point of the $\braket{\sigma_1^z(t)^2}$ curve to $1/3$. Fig:\ref{sigma1z_MBL} and \ref{sigma1z_ETH} show how the  time evolution of $\braket{\sigma_1^z(t)^2}$. 
 We define the equilibration time $T^\star$  in terms of time fluctuations. Such that for a given small standard $\epsilon = 5\times10^{-7}$, for almost all $t > T^\star$, $|\sigma_1^z(t)^2 - \overline{\sigma}_1^z(t)^2|^2 < \epsilon$. To locate the exact equilibration time, we smooth the curves and shift the equilibration value to zero, by the way, we square each data point to get  higher smoothness.
 Fig:\ref{EqTMBL} shows how local observable $\sigma_1^z$ reaches to equilibration in the MBL phase for different system sizes.  The scaling behavior of $T^\star(L)$ is shown in Fig:\ref{eqtMBLscaling}. It is the how $\log T^\star$ scales with the system size $L$. Linear fitting shows that $T^\star(L) \sim \exp L$.

\begin{figure}[h]
\centering
\includegraphics[width=1.0\linewidth]{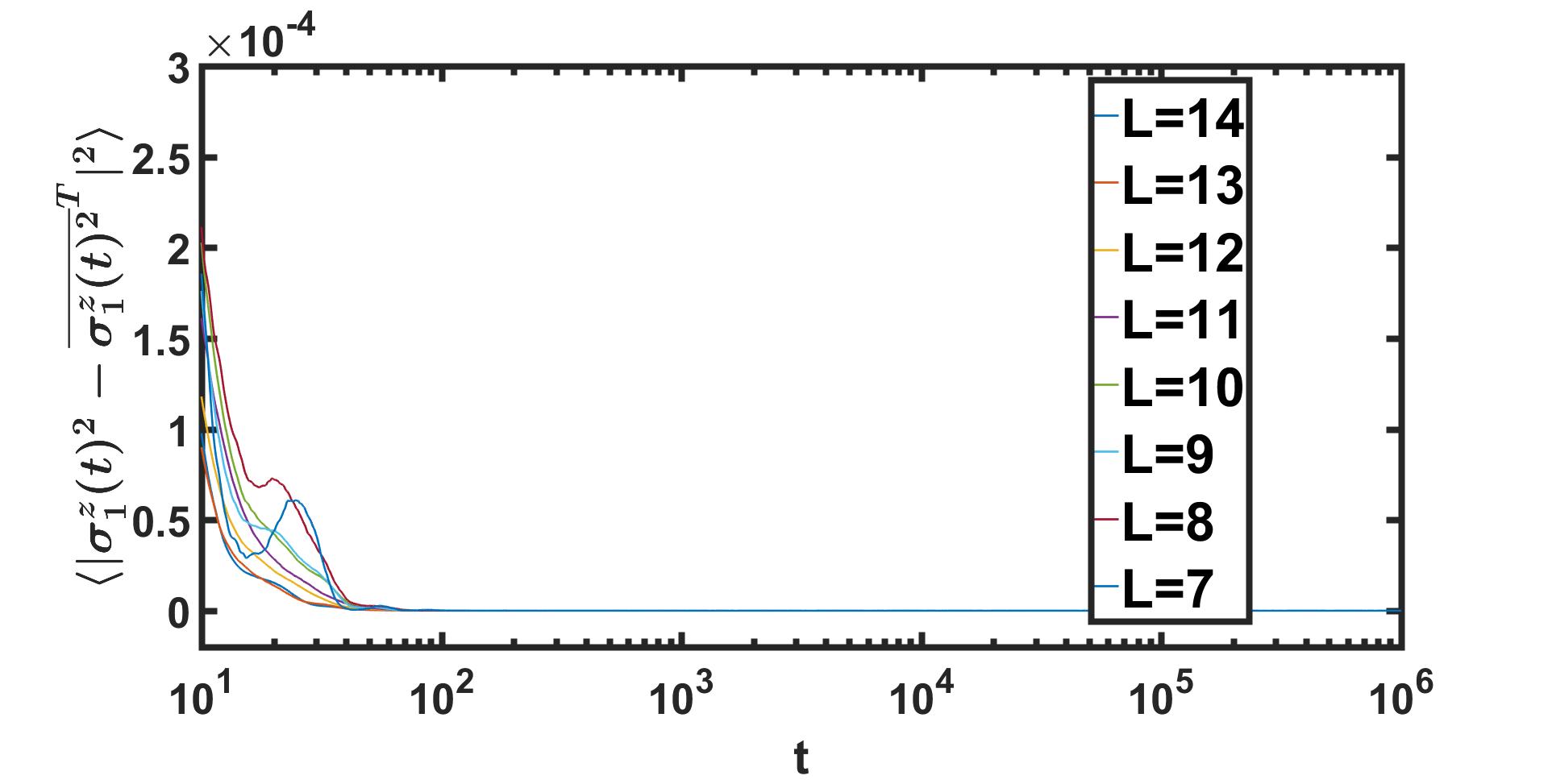}
\caption{Equilibration time scaling with the system size in the ETH phase. The system size are 7-14 respectively. For system size 7-10, each curve is averaged over 2000 realizations. For system size 11 and 12, each curve is averaged over 1000 realizations. For system size 13, it is averaged over 500 realizations. For system size 14, it is averaged over 400 realizations.}
\label{EqTETH}
\end{figure}

\begin{figure}[h]
	\centering
	\includegraphics[width=1.0\linewidth]{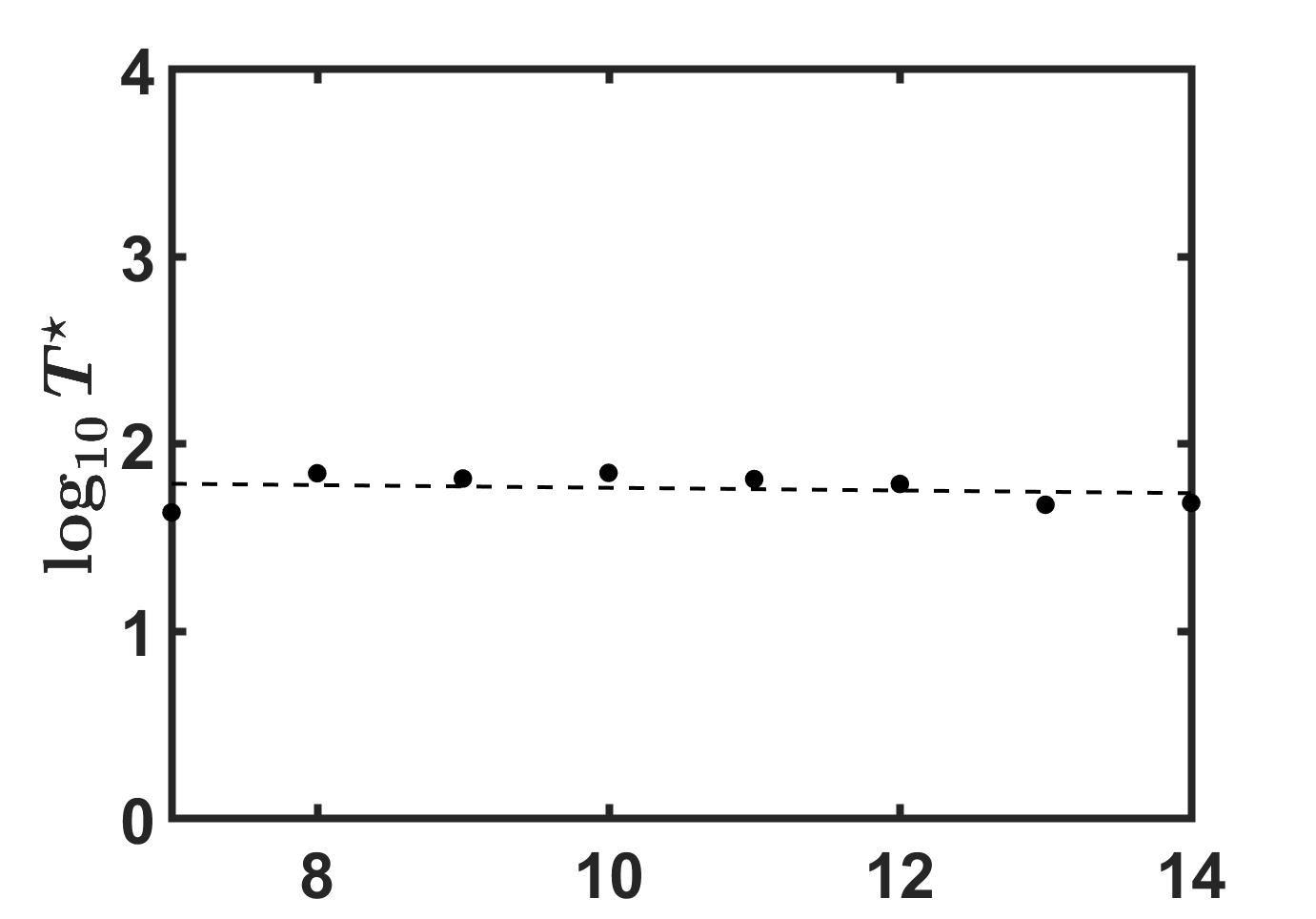}
	\caption{scaling behavior of local observable equilibration time $\log(T^\star)$ with the system size $L$ in the ETH phase with $h = 0.5$.}
	\label{eqtETHscaling}
\end{figure}

In the ETH phase, things are very different. Fig:\ref{EqTETH} shows how the $T^\star(L)$ scales with the system size. We set $\epsilon = 5\times 10^{-7}$ to locate the equilibration time. As Fig:\ref{eqtETHscaling} shows, the equilibration time does not scale with the system size $L$. Despite the difference in the system size, the equilibration time does not change, which is a constant $\sim 10^2$.

The analysis of infinite time fluctuations in terms of Loschmidt Echo and IPR is very telling in the ETH phase, because even on reasonable observation times  equilibration is reached together with all the features that one can compute from the infinite time average.  Instead, in the MBL phase, the equilibration time grows exponentially with the system size, so time fluctuations can be large in reasonable (polynomial in $L$ ) observation times. So not only the equilibrium behavior is different in terms of the scaling of the exponents $\alpha_1, \alpha_2$ between ETH and MBL. The fact that equilibrium cannot in practice be observed in the MBL phase needs to be taken in consideration.

\begin{figure}[h]
     	\centering
     	\includegraphics[width=1.0\linewidth]{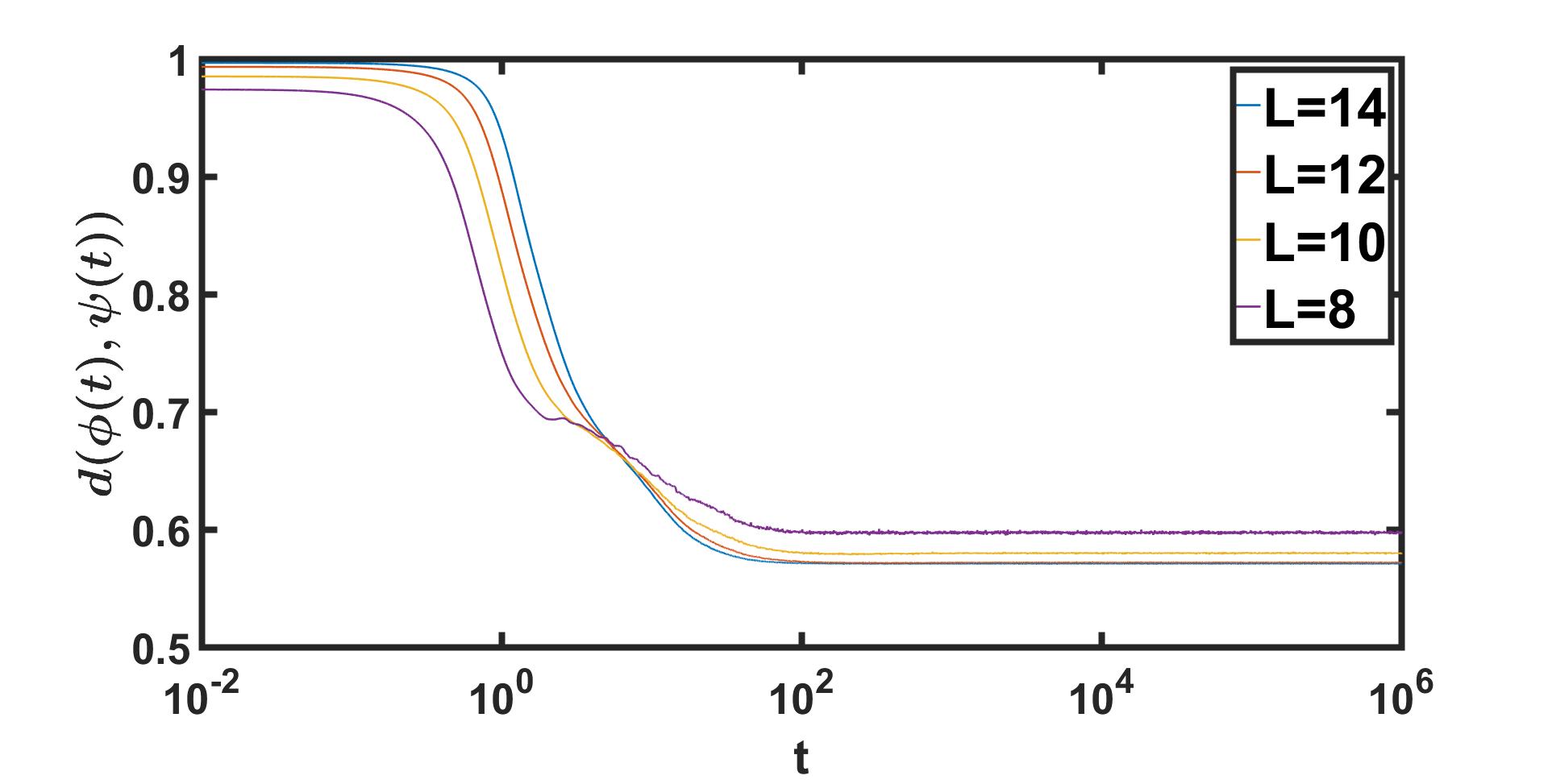}
     	\caption{Time evolution of the trace distance for the $H$ with $h=0.5$ deep in the ETH phase. the y-axial shows the trace distance, the x-axial is time. The data are averaged over 1000 realizations for $L = 8, 10, 12$, 500 realizations for $L = 14$}
     	\label{ETHscrambling}
     \end{figure}

\begin{figure}[h]
      	\centering
      	\includegraphics[width=1.0\linewidth]{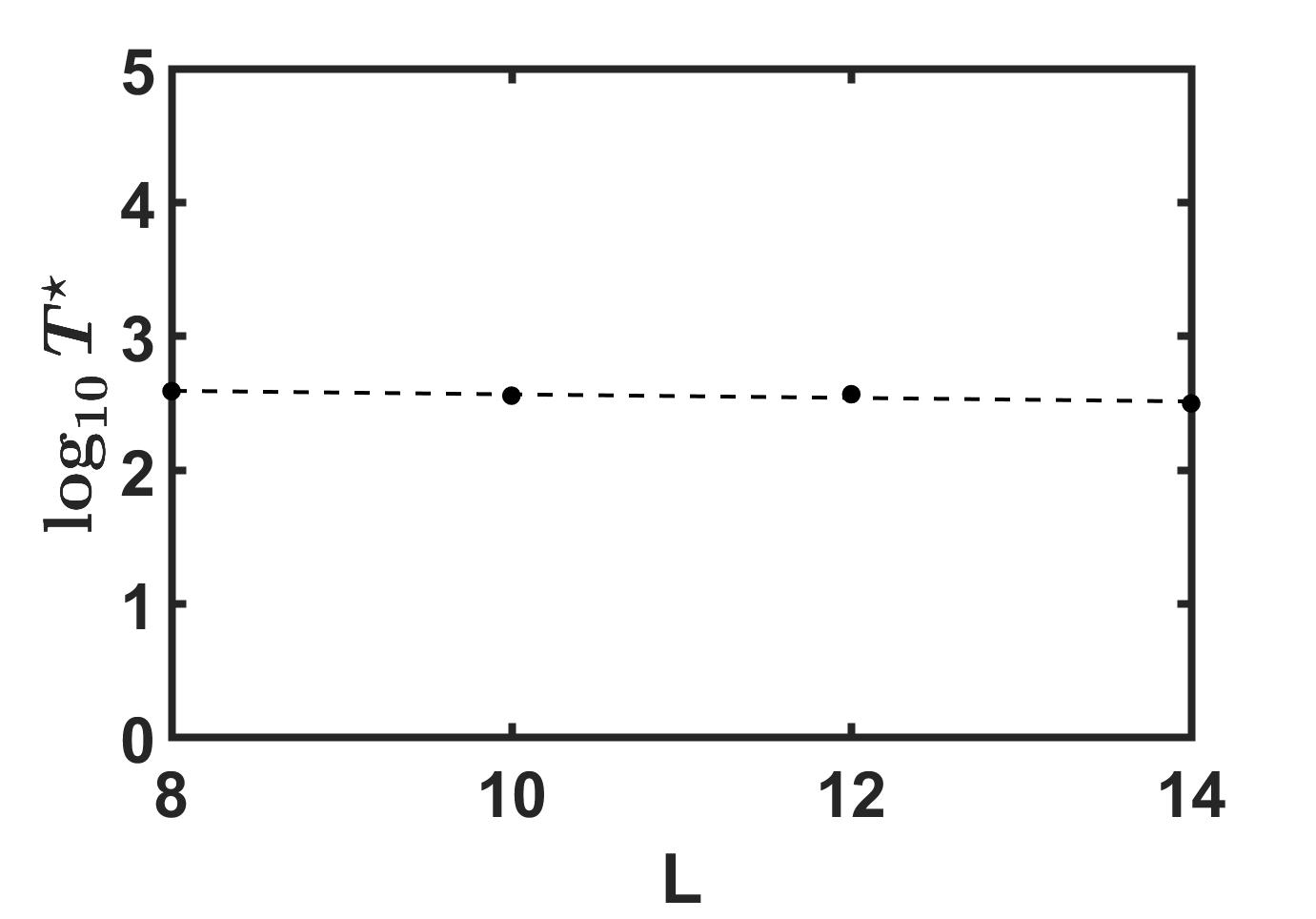}
      	\caption{logarithmic scrambling time $\log(T\star)$ scaling with the system size in the ETH phase with $h=0.5$}
      	\label{sctscalingETH}
      \end{figure}

{\em Scrambling power.---} Here, we show that the two dynamical phases are also distinguished in therms of their scrambling powers. We define scrambling as the phenomenon such that, if two states are completely distinguishable, i.e., orthogonal, they become completely undistinguishable on a subystem. The time $T^*$ for the scrambling to ensue is called scrambling time\cite{scrambling}. In this work, we use a slightly different definition of scrambling time. We define the scrambling time in terms of the time necessary to reach the maximum possible scrambling that the evolution at hand is capable of producing. 
 The scrambling protocol we used is the following:  we choose two random product states that are orthogonal to each other $\ket{\psi(0)}$ and $\ket{\phi(0)}$. We also require the two random product states to have similar energy. In our model, the energy of a state ranges approximately from $-LJ$ to $LJ$ which is of the order $O(L)$.  We choose $\delta = 0.1$ such that $|E_\phi - E_\psi| < \delta$.Then, we let
	  the two random product states evolve under our hamiltonian as  $\ket{\psi(t)} = \exp(-iHt)\ket{\psi(0)}$ and $\ket{\phi(t)} = \exp(-iHt)\ket{\phi(0)}$ and calculate the corresponding reduced density matrices $\psi_A(t) = \mathrm{Tr} _B{\ket{\psi(t)}\bra{\psi(t)}}$ and $\phi_A(t) = \mathrm{Tr} _B{\ket{\phi(t)}\bra{\phi(t)}}$, where the subsystem $A$ is taken to be half of the spin chain. 
	We measure the distinguishability of the partial states at the time $t$ by the  trace distance $d(\phi,\psi) = \frac{1}{2} \sqrt{(\phi-\psi)(\phi-\psi)^\dagger}$, averaging over $50$ realizations of initial states and disorder for a chain with   $L=14$  spins and $100$ realizations for chains with $L = 12,10,8$ spins. The scrambling time $T^\star$ is defined, given  a small enough positive $\epsilon$, as the time $T^*$ such that for almost all $t>T^\star$, $|d(\phi(t),\psi(t)) - \overline{d(\phi(t),\psi(t)}| < \epsilon = 5\times 10^{-7} $.
In the ETH phase, see Fig.\ref{ETHscrambling} and Fig.\ref{sctscalingETH}, we see that   the logarithmic  scrambling time is upper bounded by a constant. 
   
  The MBL phase behaves dynamically in a dramatically different way also in terms of its scrambling power. Let us now see the time evolution of the trace distance and the finite size scaling      of the scrambling time in the MBL phase. As shown in Fig:\ref{sctscalingMBL}, the logarithm of scrambling time  behaves like $\log  T^\star \propto L$ in the MBL phase.
  
     \begin{figure}[h]
   	\centering
   	\includegraphics[width=1.0\linewidth]{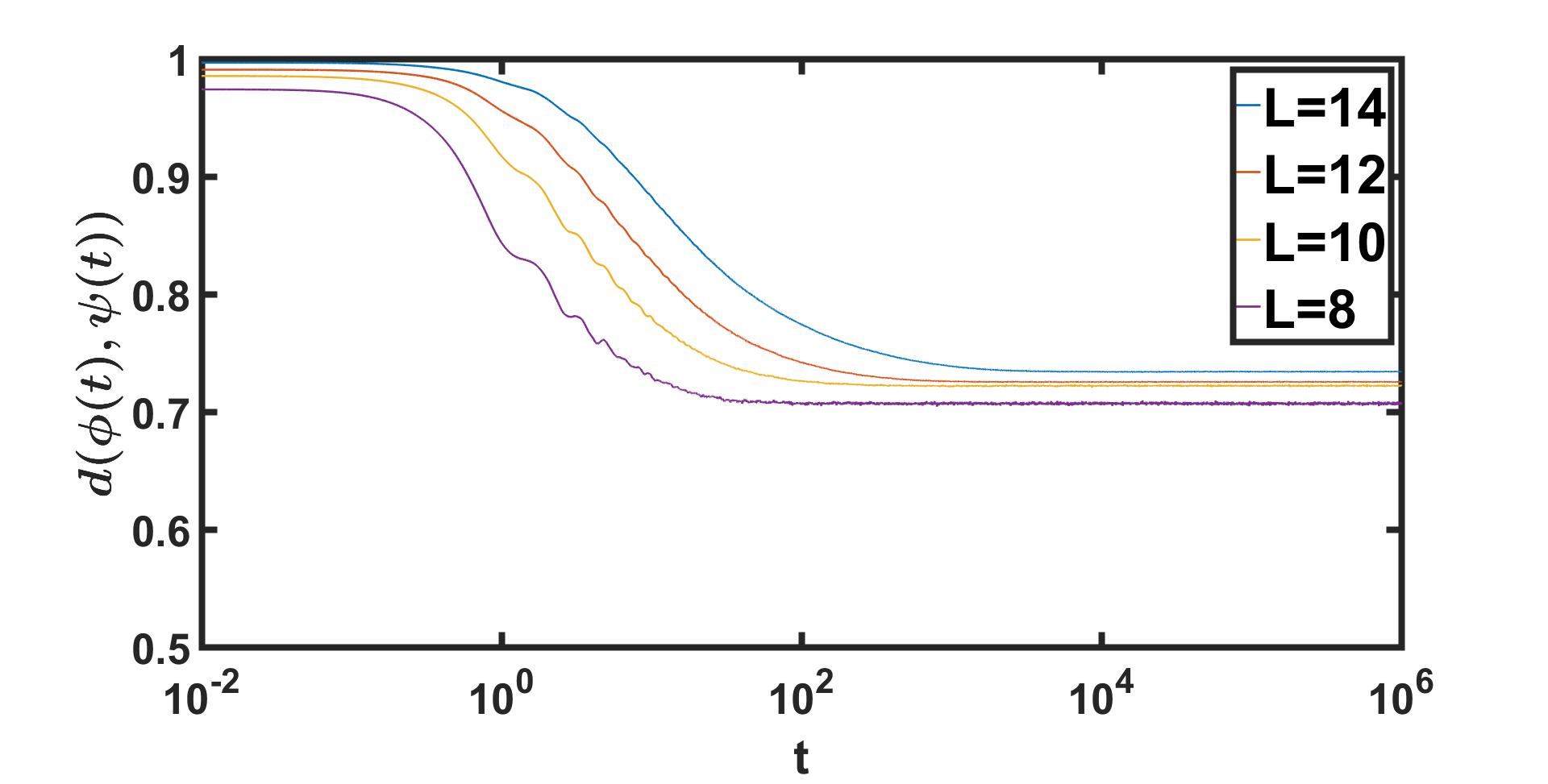}
   	\caption{Time evolution of the trace distance $h=6.5$, the y-axial shows the trace distance, the x-axial is time. The data are averaged over 1000 realizations for $L = 8, 10, 12$, 500 realizations for $L = 14$}
   	\label{MBLscrambling}
   \end{figure}

   \begin{figure}[h]
   	\centering
   	\includegraphics[width=1.0\linewidth]{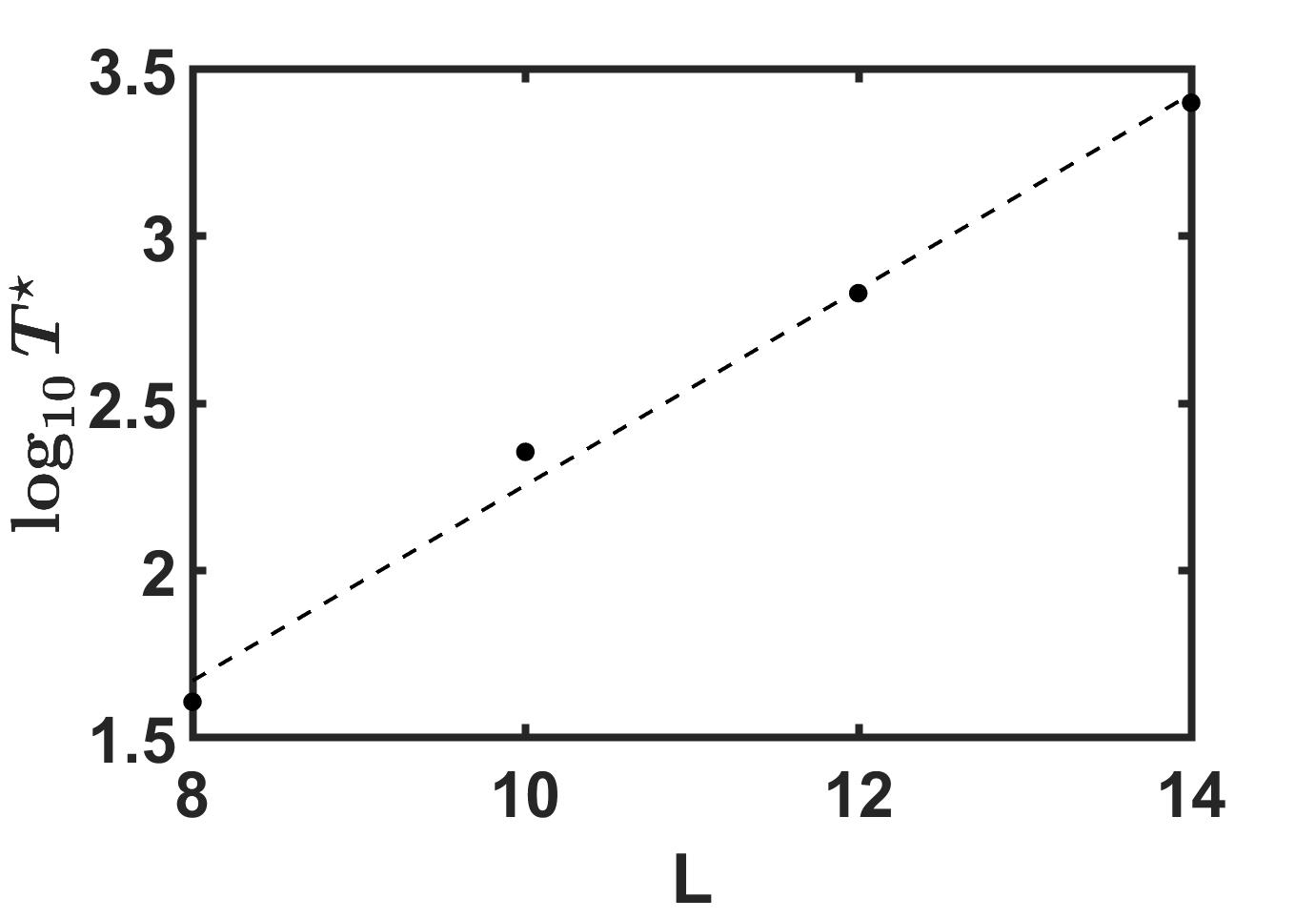}
   	\caption{logarithmic scrambling time $\log(T^\star)$ scaling with the system size in the MBL phase with $h=6.5$}
   	\label{sctscalingMBL}
   \end{figure}

{\em Conclusions.---} In this paper, we study the dynamics after a quantum quench in both the ETH and the MBL phase. The dynamical behaviour is investigated using the Loschmidt Echo and its temporal fluctuations on both the full system and in a subsystem. Both methods give a critical point for the MBL transition around $h_c\sim 3 \pm .3$, which is compatible with other known estimates. We remark that, given the fact that the system is relaxing slowly in most of its local observables, the infinite time average of the Loschmidt Echo does not characterize the size of fluctuations in reasonable times. As it was very recently shown, these do damp as a power law instead, which is obtained by computing the Loschmidt Echo on finite time scales, see \cite{serbyn}
 However, it is remarkable that the scaling of such temporal fluctuations even in the infinite time limit serves as a order parameter for the MBL transition. Finally, we show that the scrambling power of the MBL phase is dramatically different from that of the ETH phase, as result of its slow dynamics.



{\em Acknowledgments.---} 
This work was supported in part by the National Basic Research Program of China Grant 2011CBA00300, 2011CBA00301 the National Natural Science Foundation of China Grant  61033001, 61361136003,11574176 


\section{APPENDIX}
\subsection{proof of the statement that $\braket{\sigma_1^z(0)^2} = 1/3$}
The initial random product states are generated with the following steps\cite{Tth2008430}. 
\begin{enumerate}
	\item Choose two random variables $\alpha, \beta \in \mathbb{C}$ whose real and imaginary part satisfy Gaussian distribution.
	\item Construct a normalized vector $$\frac{1}{\sqrt{|\alpha|^2+|\beta|^2}}\left(\begin{matrix}
	\alpha
	\\ 
	\beta
	\end{matrix} \right)$$
	\item Tensor product $L$ times for system size equals $L$.
\end{enumerate}

Thus 
\begin{equation}
	\ket{\psi(0)} = \otimes_{i=1}^L (\alpha_i \ket{0} + \beta_i \ket{1})
\end{equation}

where $|\alpha_i|^2+|\beta_i|^2 = 1$
\begin{equation}
	\sigma_1^z(0)=\braket{\psi(0)|\sigma_1^z|\psi(0)} = |\alpha_1|^2 - |\beta_1|^2
\end{equation}
	
\begin{equation}
	\braket{\sigma_1^z(0)^2} = \braket{(|\alpha_1|^2 - |\beta_1|^2)^2} = \braket{4|\alpha_1|^4-4|\alpha_1|^2+1}
\end{equation}

In the following part we calculate the value of $\braket{|\alpha_1|^2}$ and $\braket{|\alpha_1|^4}$

We write $\alpha_1 $ and $\beta_1$ as $x_1+i y_1$ and $x_2 + i y_2$
\begin{align*}
	\braket{|\alpha_1|^2} =&
	 \int_{-\infty}^\infty \mathrm{d}x_1\mathrm{d}x_2\mathrm{d}y_1\mathrm{d}y_2 \frac{x_1^2+y_1^2}{x_1^2+y_1^2+x_2^2+y_2^2}p(x_1)\\
	 &p(x_2)p(y_1)p(y_2) = \frac{1}{2}
\end{align*}

\begin{align*}
\braket{|\alpha_1|^4} =&
\int_{-\infty}^\infty \mathrm{d}x_1\mathrm{d}x_2\mathrm{d}y_1\mathrm{d}y_2 \frac{(x_1^2+y_1^2)^2}{(x_1^2+y_1^2+x_2^2+y_2^2)^2}p(x_1)\\
&p(x_2)p(y_1)p(y_2) = \frac{1}{3}
\end{align*}

Where $p(x)=\frac{1}{\sqrt{2\pi}\sigma}\exp(-\frac{x^2}{2\sigma^2})$

Thus $\braket{\sigma_1^z(0)^2} = 1/3$.
\subsection{proof of the statement that $\braket{\mathcal{L}}\leq (2/3)^L$ with random product initial states}

$\ket{k}_c$ denotes the computational basis, $\ket{n}_E$ is the energy eigenstate.

We write the energy eigenstate as the linear combination of computational basis.
\begin{equation}
	\ket{n}_E = C^{(n)}_k \ket{k}_c
\end{equation}
where $\sum_{k=1}^{2^L} |C^{(n)}_k|^2 = 1$.

\begin{equation}
	\ket{\psi(0)} = \alpha_1\alpha_2\dots\alpha_L\ket{00\dots0}+\dots=\sum_{k=1}^{2^L} \mathcal{P}_k\ket{k}_c
\end{equation}

Thus
\begin{equation}
	\begin{split}
	&\braket{\mathcal{L}} = \sum_{n = 1}^{2^L} |\braket{n|\psi(0)}|^4 \\
	&= \sum_{n = 1}^{2^L} \sum_{r,s,t,q=1}^{2^L} \braket{C^{(n)\star}_rC^{(n)\star}_sC^{(n)}_tC^{(n)}_q} \braket{\mathcal{P}_r\mathcal{P}_s\mathcal{P^\star}_t\mathcal{P^\star}_q}
	\end{split}
\end{equation}

The average

 $\braket{C^{(n)}_r} = 0, \braket{\mathcal{P}_k} =0, \braket{|\mathcal{P}_k|^2} = (1/2)^L, \braket{|\mathcal{P}_k|^4} = (1/3)^L$.
 
 Each term is non-vanishing if and only if $r=s,t=q$. Thus

 \begin{equation}
 \begin{split}
 	\braket{\mathcal{L}} &= \sum_{n = 1}^{2^L}\left(\sum_{r = 1}^{2^L} |C^{(n)}_r|^4\right)\left(\frac{1}{3}\right)^L +
 	\sum_{n = 1}^{2^L}\left(\sum_{r \neq t}^{2^L} |C^{(n)}_r|^2|C^{(n)}_t|^2\right)\left(\frac{1}{4}\right)^L\\ &= 
 	\sum_{n = 1}^{2^L}\left(\sum_{r = 1}^{2^L} |C^{(n)}_r|^4\right)\left(\frac{1}{3}\right)^L + 
 	\sum_{n = 1}^{2^L}\left(1-\sum_s |C^{(n)}_s|^4\right)\left(\frac{1}{4}\right)^L \\
 	&\leq 2^L\left[(\frac{1}{3})^L - (\frac{1}{4})^L\right] + 2^L(\frac{1}{4})^L = (2/3)^L
 \end{split}
 \end{equation}


\end{document}